\documentclass[11pt]{article}
\usepackage{moriond,epsfig}

\def\be{\begin{equation}}
\def\ee{\end{equation}}
\def\bea{\begin{eqnarray}}
\def\eea{\end{eqnarray}}

\begin{document}
\vspace*{4cm}
\title{RECENT RESULTS FROM THE KLOE EXPERIMENT AT DAPHNE}

\author{ THE KLOE COLLABORATION\footnote{The KLOE Collaboration:
A.~Aloisio,
F.~Ambrosino,
A.~Antonelli,
M.~Antonelli, 
C.~Bacci,
G.~Bencivenni, 
S.~Bertolucci, 
C.~Bini, 
C.~Bloise, 
V.~Bocci
% ...
G.~Xu.
 } \\ \ \ \   \\
         Presented by ANTONIO PASSERI }

\address{INFN Sezione Roma III and Dipartimento di Fisica, Universit\'a Roma Tre, via della Vasca Navale 84, 00146 Roma, Italy}

\maketitle\abstracts{
Recent results from the KLOE experiment at the DA$\Phi$NE $e^+e^-$ collider 
are presented. KLOE has collected $\sim\, 500 pb^{-1}$ of data in the years 
from 2000 to 2002. Preliminary results are obtained using wide samples of the 
full statistics and include: the BR of the $K_{e3}$ decay of the $K_S$ and the
first measurement of its charge asymmetry, the ratio 
BR($K_S\rightarrow \pi^+\pi^-(\gamma)$)/BR($K_S\rightarrow \pi^0\pi^0$), the ratio 
BR($K_L\rightarrow \gamma\gamma$)/BR($K_L\rightarrow 3\pi^0$), a detailed study
of the decay $\phi\rightarrow\pi^+\pi^-\pi^0$, the ratio 
BR($\phi\rightarrow \eta^{\prime}\gamma$)/BR($\phi\rightarrow \eta\gamma$) and
the pseudoscalar mixing angle $\varphi_P$, the $\phi\rightarrow f_0\gamma$ and 
the $\phi\rightarrow a_0\gamma$ BR's. 
}

\section{Introduction}
DA$\Phi$NE \cite{dafne} is an $e^+e^-$ collider working at the $\phi$ resonance
peak, and located in the Frascati INFN laboratories. The $\phi$ meson is 
produced essentially at rest and decays abundantly ($\sim 34\%$) to $K_SK_L$, 
which is a pure $J^{PC}=1^{--}$ quantum state. Thus DA$\Phi$NE provides two 
highly pure, almost monochromatic, back-to-back $K_S$ and $K_L$ beams.\par
The KLOE experiment \cite{kloe} is designed to exploit the unique features of 
a $\phi$-factory environment for the measurement of CP and CPT violation in 
the $K^0-\bar K^0$ system, and more generally for the study of kaons' decays. 
In particular, KLOE has the unique capability of 
tagging the presence of a $K_S$ ($K_L$) by detecting a $K_L$ ($K_S$) flying 
in the opposite direction. $K_S$ and $K_L$ are easily distinguishable on the 
basis of their lifetimes: due to the small boost, $K_S$'s have a mean decay 
path $\lambda_S\sim 6 \, mm$ and they all decay near the interaction point, 
while $\lambda_L\sim 340\, cm$. Other abundant $\phi$ decay modes are 
$K^+K^-$ ($\sim 49\%$), for which a tagging technique can be applied similar 
to the neutral kaon case, and $\rho\pi$ ($\sim 15\%$).\par
The DA$\Phi$NE collider has been commissioned in 1999, when it delivered its 
first 2 pb${}^{-1}$ of data, and it has been operated with continously 
improving luminosity up to 2002, when it reached a maximum peak luminosity of 
7.8$\cdot 10^{31}cm^{-2}s^{-1}$. The total integrated luminosity collected by 
KLOE amounts now to 500 pb${}^{-1}$.
The 25 pb${}^{-1}$ year 2000 data sample has been fully analized and results have been 
recently published \cite{kspienu,ratio,eta,f0,a0}. Year 2001 (190 pb${}^{-1}$)
and 2002 (300 pb${}^{-1}$) data 
samples are now being analyzed and some preliminary results are presented here.\par

\section{The KLOE detector}
\begin{figure}[bt]
\begin{center}
\epsfig{figure=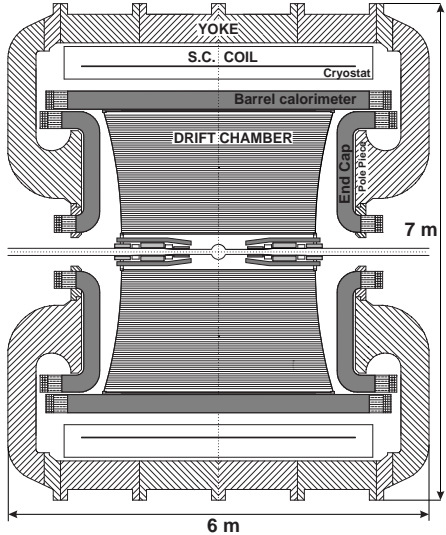,width=6cm}
\end{center}
\caption{Schematic view of the KLOE detector.
\label{fig:kloeview}}
\end{figure}

A schematic side view of the KLOE detector is shown in fig.\ref{fig:kloeview}.
It consists mainly in a large volume drift chamber surrounded by an 
electromagnetic calorimeter. A superconductiong coil provides a 0.52 T solenoidal magnetic field.\par
The calorimeter\cite{calo} is a fine sampling lead-scintillating fiber one, 
composed by 24 barrel modules and 2x32 end modules,
with photomultiplier readout at both sides. Module thickness is 23 cm ($\sim
15X_0$) and the total solid angle coverage is 98\% . The energy resolution, as 
measured using $e^+e^-\rightarrow e^+e^-\gamma $ events, is $\sigma_E / E =
5.7\% /\sqrt{E(GeV)} $. The intrinsic time resolution, as measured using  
 $e^+e^-\rightarrow e^+e^-\gamma $ and  $e^+e^-\rightarrow 2\gamma $ events,
is $\sigma_t = 54\, ps/\sqrt{E(GeV)} \oplus 50\, ps$. Two smaller calorimeters,
QCAL\cite{qcal}, made with lead and scintillating tiles are wrapped around 
the low beta quadrupoles to complete the hermeticity.\par
The drift chamber\cite{dc} is a cylinder, 3.3 m long and having 4 m diameter, strung 
with $\sim$ 52000 wires (of which $\sim$ 12000 are sense wires) in an all 
stereo configuration. In order to minimize multiple scattering and $K_L$ 
regeneration and to maximize detection efficiency of low energy photons, 
the chamber works with a 90\% helium - 10\% isobutane gas mixture, while its 
walls are made of light materials (mostly carbon fiber composites). The 
momentum resolution for tracks at large polar angle is $\sigma_p / p
\leq 0.4\%$,  the spatial resolutions are $\sigma_{r,\phi}\approx\, 
150\mu m$ and $\sigma_z\approx\, 2mm$. \par

\section{$K_S$ decays}
The presence of a $K_S$ is tagged by identifying the interactions of the $K_L$
in the calorimeter. Approximately 30\% of the $K_L$'s produced in 
$\phi\rightarrow K_SK_L$ events reaches the calorimeter before decaying and 
interacts therein. Such an interaction, referred to as {\it $K_L$ crash}, has
a very clear signature consisting of a high energy ($E>200\, MeV$) cluster in 
the calorimeter, which is also neutral (i.e. not associable to any track in the
event) and delayed ($\sim 30 \, ns$ after all other clusters in the event, due
to $K_L$ small velocity, $\beta \sim 0.22$). The cluster position,
exploiting the $\phi$ decay kinematics, provides the momentum of the $K_S$.
Moreover about 40\% of the $K_L$ crashes independently satisfy the trigger 
conditions, thus facilitating trigger efficiency studies.\par 

\subsection{BR($K_S\rightarrow \pi^+\pi^-(\gamma)$)/BR($K_S\rightarrow \pi^0\pi^0$)}
This ratio is the first part of the double ratio used to measure 
$\Re (\epsilon^{\prime}/\epsilon)$.\par
$K_S\rightarrow\pi^+\pi^-$ decays are identified, in the sample tagged by $K_L$
crash, by requiring the presence of two oppositely charged tracks, originating 
from the interaction region and satisfying loose cuts in momentum $p$ and 
polar angle $\theta$. Such cuts define the acceptance for the decay, which is 
evaluated using Monte Carlo simulation (MC), while the single track 
reconstruction efficiency is evaluated directly from subsamples of $K_S
\rightarrow\pi^+\pi^-$ events, in $(p,\theta )$ bins. The ratio of data and MC 
efficiency is found to be constant over all the acceptance region, and MC efficiency is scaled accordingly. 
The previous selection includes also $K_S\rightarrow\pi^+\pi^-\gamma$ events,
with an efficiency decreasing with increasing $E_{\gamma}$ (or with decreasing
$\pi^+\pi^-$ invariant mass). Such efficiency is obtained from simulation down
to $20\, MeV$ photons, extrapolated to $E_{\gamma}=0$ and folded with the 
theoretical photon spectrum \cite{cirigliano} to obtain the overall efficiency.
No threshold is applied to $E_{\gamma}$.\par 
$K_S\rightarrow\pi^0\pi^0$ decays are selected (still in the $K_S$ tagged 
sample) by requiring the presence of at least 3 prompt neutral clusters
in the calorimeter. A cluster is defined to be prompt when $| T_{cl}-R/c | <
5\sigma_t $, where $T_{cl}$ is the cluster time of flight, $R$ is the calorimeter
radius and $\sigma_t$ is the measured time resolution. Loose cuts in energy and
polar angle are applied to each cluster, thus defining the acceptance for the
decay, which is then obtained from simulation. The photon detection efficiency
is evaluated using $\phi\rightarrow\pi^+\pi^-\pi^0$ events, in which one photon
and the two pions tracks are used to determine the kinematics of
the remaining photon, which is then searched for.\par
The trigger efficiency is evaluated as the combined probability of the $K_S$ 
decay and of the $K_L$-crash to satisfy the trigger condition. Events in which
the $K_L$-crash alone is enough to satisfy the trigger are used to estimate the
trigger probability of the $K_S$ decay, and vice versa. Using only a 17 
pb${}^{-1}$ data sample, collected in year 2000, the final result 
is obtained\cite{ratio} 
\begin{equation} 
\frac{\Gamma (K_S\rightarrow\pi^+\pi^-(\gamma))} 
{\Gamma (K_S\rightarrow\pi^0\pi^0)}  = 2.239\pm 0.003_{stat} \pm 0.015_{syst}.
\label{eq:ratio} 
\end{equation} 
This measurement is compatible with the world average \cite{pdg} at 
2.7$\sigma$ level, and it has the best statistical significance ever reached. 
It is also the first that takes fully into account the photon radiation.
The ratio in (\ref{eq:ratio}) can be expressed in terms of the transition
amplitudes between states of $\Delta I=0$ ($A_0$) and $\Delta I=2$ ($A_2$),
and of their phase shift $\delta_2 - \delta_0$. Such phase
shift can be predicted from chiral 
perturbation theory\cite{gasser1} to be $(45\pm 6)^{\circ}$, 
while from $\pi\pi$ scattering\cite{gasser2} it is estimated to be 
$(47.7\pm 1.5)^{\circ}$. If then it is assumed $(A_0/A_2)^2=
(3\tau_S /4\tau)BR(K^{\pm}\rightarrow\pi^{\pm}\pi^0)^{-1} -1= (22.2\pm0.07)^2$,
the PDG value for $\Gamma (K_S\rightarrow\pi^+\pi^-)/\Gamma (K_S\rightarrow
\pi^0\pi^0)$ yields $\delta_2 - \delta_0=(56.7\pm 3.8)^{\circ}$, while 
using the present KLOE measurement
we obtain $\delta_2 - \delta_0=(48\pm 3)^{\circ}$, in good agreement with predictions.\par
  
\subsection{BR($K_S\rightarrow\pi^{\pm}e^{\mp}\bar\nu (\nu)$)}
Assuming CPT conservation and the $\Delta S=\Delta Q$ rule, the $K_S$ and $K_L$
partial widths for the $K_{e3}$ decay must be equal\cite{maiani}, and the corresponding 
$K_S$ branching ratio can be easily obtained from that of the $K_L$. KLOE has 
performed a direct measurement of this $K_S$ branching ratio\cite{kspienu} 
using the year 2000 data sample. The update of such measurement with 2001 data
is presented here.\par 
$K_S\rightarrow\pi e \nu$ events are selected, in the $K_L$-crash tagged 
sample, by requiring the presence of two oppositely charged 
tracks which extrapolate and form a vertex in the interaction region. Loose
momentum and angular cuts are applied, and the vertex invariant mass, evaluated
in the hypothesis that both tracks belong to pions, is required to be smaller
than 490 $MeV$, thus rejecting 95\% of the $K_S\rightarrow\pi^+\pi^-$ decays.
Vertex reconstruction and preselection efficiencies are evaluated by MC.\par
\begin{figure}[tb]
\begin{center}
\begin{tabular}{ll}
\epsfig{figure=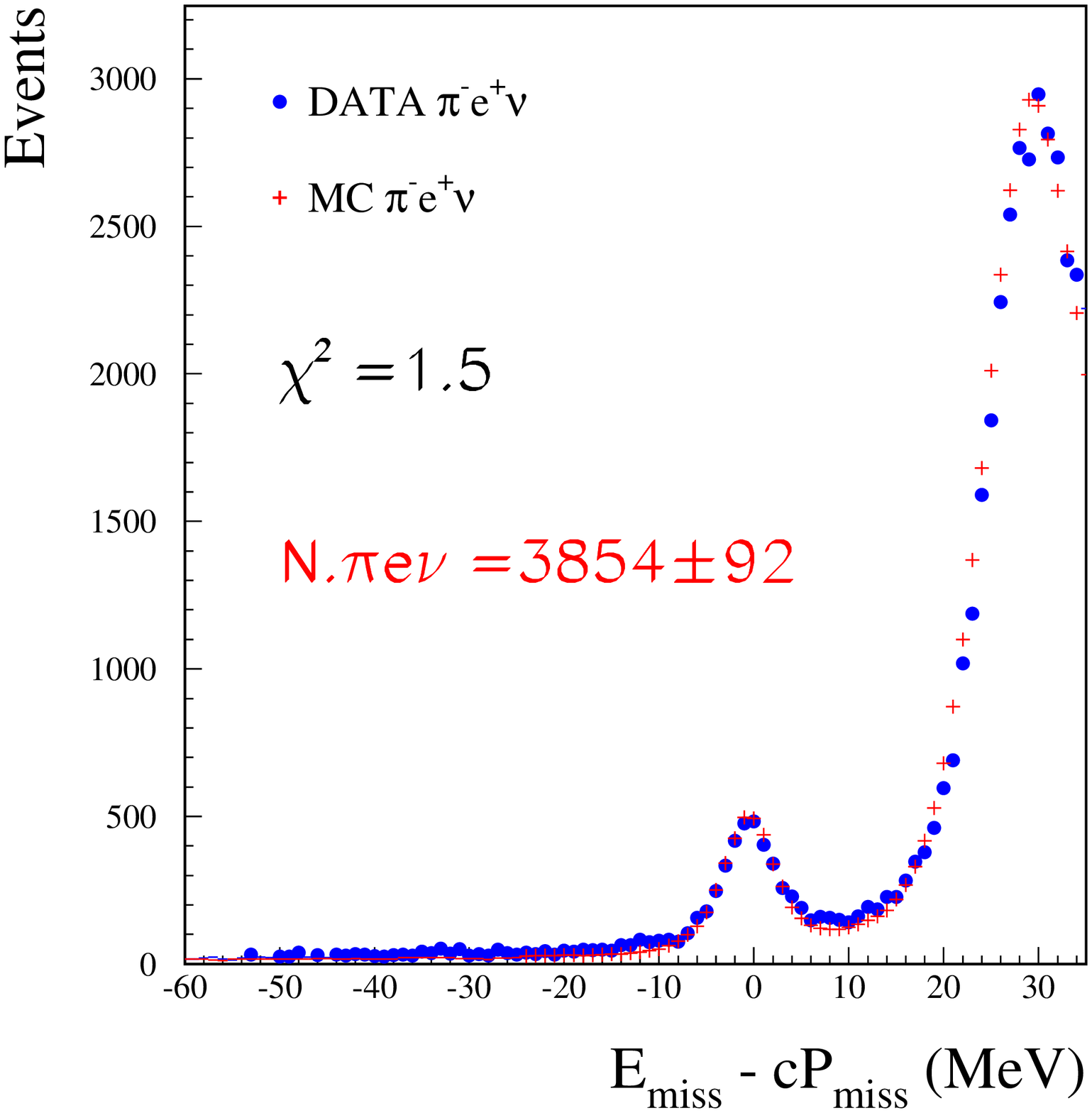,width=8cm} & \epsfig{figure=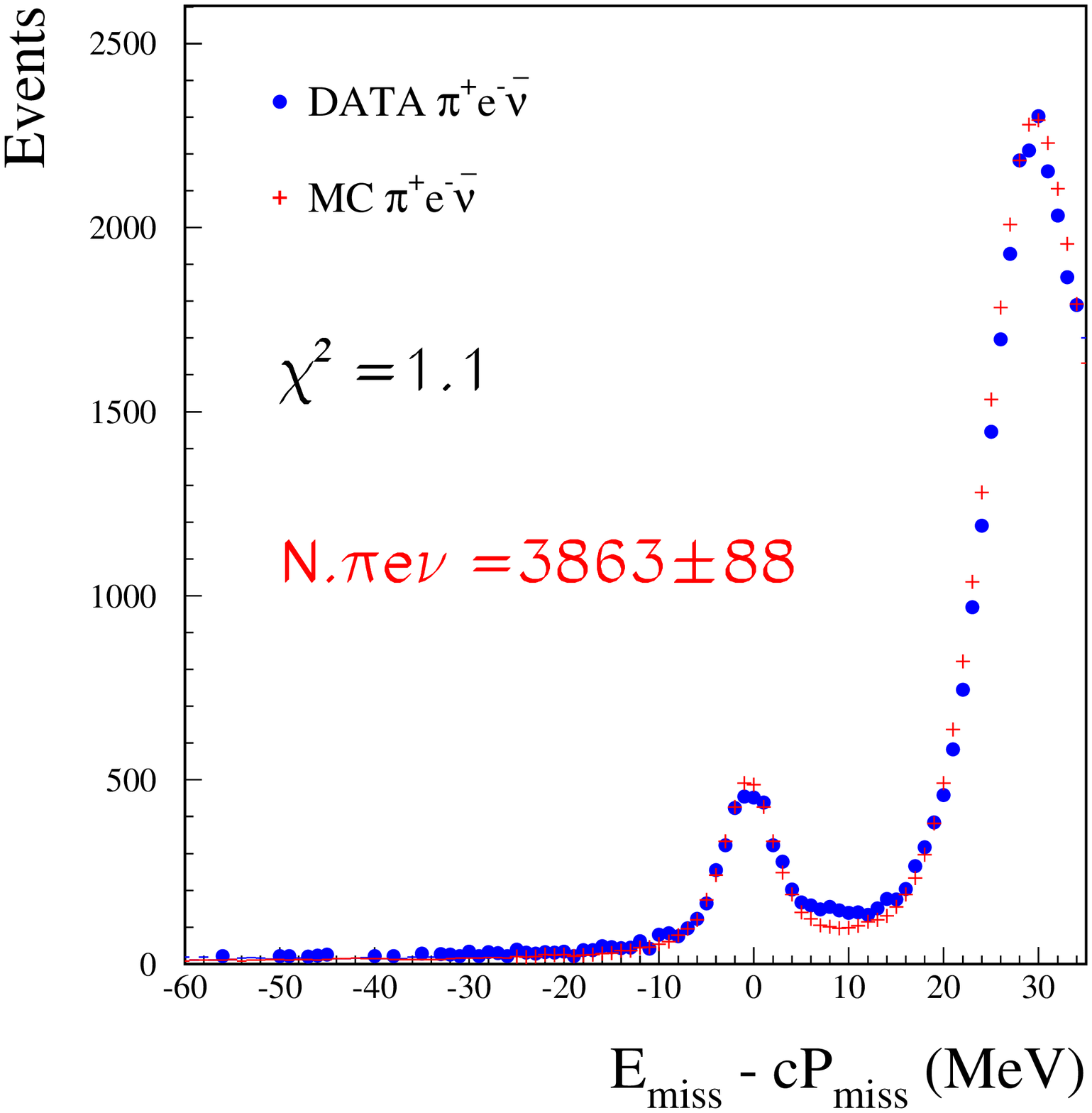,width=8cm} \\
\end{tabular}
\end{center}
\caption{$E_{miss}-P_{miss}$ distribution for  $K_S\rightarrow\pi e \nu$ 
candidates. Solid markers represent data, crosses are the MC based fit.
\label{fig:kspienu}}
\end{figure}
The $K_S\rightarrow\pi e \nu$ decays are identified and the $\pi / e$ 
assignement is made by means of the time of flight of the two tracks, which 
are then both required to be associated to a calorimeter cluster. For each 
track the difference $\delta_t(m)=t_{cl}-L/\beta(m)$ is computed, where 
$t_{cl}$ is the time of the associated cluster, $L$ is the measured track 
length and $\beta(m)$ is the particle velocity obtained from the measured
momentum with a mass hypotesis $m=m_e$ or $m=m_{\pi}$. In order to avoid
any systematics related to the $T_0$ determination, the selection is made on
the difference $\delta_{t,ab}=\delta_t(m_a)_1 - \delta_t(m_b)_2$, where mass
hypothesis $a$ ($b$) is used for track 1 (2). For the correct mass assignement
we expect $\delta_{t,ab}\sim 0$, therefore a cut is applied at $\delta{t,ab}<1
\, ns$. The $\pi /e$ identification efficiency is measured in a sample of $K_L
\rightarrow\pi e \nu$ in which the $K_L$ decays near the interaction region.
Such a sample can be selected with high purity (99.7\%) just by means of 
kinematics, without using calorimeter times. The track-to-cluster association
efficiency is also measured in the same $K_L$ decays sample, in a $K_S
\rightarrow\pi^+\pi^-$ sample and in a $\phi\rightarrow\pi^+\pi^-\pi^0$ 
sample. In all such topologies the events are reconstructed independently on
the presence or absence of a calorimeter cluster from one of the tracks.
The overall efficiency to detect a $K_S\rightarrow\pi e \nu$ decays is 
$0.208\pm 0.004$.\par
In fig.\ref{fig:kspienu} it is shown the distribution of the selected events in
$E_{miss}-P_{miss}$ (the neutrino energy and momentum, if the mass assignement
is correct) for the two possible charge states, obtained using 170 pb${}^{-1}$
collected in 2001 (solid markers). Signal events are included in the peak around zero, while 
the residual $K_S\rightarrow\pi^+\pi^-$ background shows up in the positive
region. The number of signal events in the distribution is obtained from a 
fit, shown in the figure with crosses, which uses the MC distributions for 
signal and background having as a free parameter their independent 
normalizations. The fit yelds $N_{\pi^+e^-\bar\nu} = 3863\pm 88$, 
$N_{\pi^-e^+\nu} = 3854\pm 92$, and from the sum of the two distributions:
$N_{\pi^{\pm}e^{\mp}\nu(\bar\nu)} = 7732\pm 127$. By normalizing these numbers
to the number of $K_S\rightarrow\pi^+\pi^-$ in the same data sets we get the preliminary KLOE results: 
\begin{equation} 
\begin{array}{ll}
BR(K_S\rightarrow\pi^+e^-\bar\nu)& = (3.44\pm 0.09\pm 0.06)\cdot 10^{-4} \\
BR(K_S\rightarrow\pi^-e^+\nu)& = (3.31\pm 0.08\pm 0.05)\cdot 10^{-4} \\
BR(K_S\rightarrow\pi^{\pm}e^{\mp}\nu(\bar\nu))& = (6.76\pm 0.12\pm 0.10)\cdot 10^{-4} \\
\end{array}
\label{eq:kspienu} 
\end{equation} 

On the basis of such results it is possible to build the $K_S$ semileptonic 
charge asimmetry $A_S = (\Gamma^+_S -\Gamma^-_S)/ (\Gamma ^+_S +\Gamma^-_S)$,
where $\Gamma^{+(-)}_S$ are the partial decay widths of the $K_S$ into
$\pi^{+(-)}e^{-(+)}\nu$. If CPT invariance holds, $A_S$ must be equal to the
corresponding $K_L$ asimmetry $A_L$\cite{maiani}. KLOE has performed the first 
measurement of $A_S$ ever done, obtaining the following preliminary result
\begin{equation} 
A_S = (1.9\pm 1.7_{stat} \pm 0.6_{syst})\cdot 10^{-2}
\label{eq:ksasimm} 
\end{equation} 
which is compatible with the present (much more precise) $A_L$ measurement
\cite{ktev}. It is also possible to test the $\Delta S=\Delta Q$ rule, assuming
the validity of CPT invariance. The relevant parameter for such test is 
$\Re (x^+)\approx \langle \pi^- e^+\nu |H_{weak}|\bar K^0\rangle /
  \langle \pi^- e^+\nu |H_{weak}| K^0\rangle  $\cite{maiani}, which can be 
extracted from the $K_S$ and $K_L$ semileptonic decay partial amplitudes: 
$\Gamma^{semil}_S /\Gamma^{semil}_L = 1+4\Re (x^+)$. The present KLOE 
measurement of $BR(K_S\rightarrow \pi e \nu)$ yields $\Re (x^+) = (2.2\pm 5.3
\pm 3.5)\cdot 10^{-3}$, in good agrement with the previous measurement 
performed by CPLEAR\cite{cplear}.\par
   
\section{$K_L$ decays}
The presence of a $K_L$ is tagged by identifying the decay $K_S\rightarrow
\pi^+\pi^-$. Such events are easily selected by requiring two oppositely 
charged tracks which form a vertex laying  in cylindrical fiducial volume 
centered in the interaction region, having radius of $4\, cm$ and length of
$16\, cm$. No other tracks must be connected to the vertex, and loose cuts are 
applied to its total momentum and invariant mass. The measured $K_S$ momentum
provides a very good estimate of the $K_L$ one, with an angular resolution
of $\sim 1^{\circ}$ and a momentum resolution of $\sim 2\, MeV$. The overall
tag efficiency is about 85\%.\par
\subsection{BR($K_L\rightarrow \gamma\gamma$)/BR($K_L\rightarrow 3\pi^0$)}
A measurement of the $K_L\rightarrow\gamma\gamma$ decay rate provides 
interesting tests of chiral perturbation theory and can be used as a constraint
for the calculation of the $K_L\rightarrow\mu^+\mu^-$ decay rate. KLOE has 
performed such measurement using a sample of 362 pb${}^{-1}$ collected during 
2001 and 2002\cite{kl2gam}. $K_L$ decays are searched for in a fiducial volume defined by
$30\, cm < r_t < 170\, cm$ ($r_t$ is the transverse radius) and $|z|<140\, cm$.
About 31.5\% of the $K_L$ decay in such a volume. The position of the $K_L$ 
neutral decay vertex can be determined by each of the photons emitted in the
decay, using the flight direction of the $K_L$ (provided by the $K_S$) and the
position and time of the photon cluster on the calorimeter. The final value of 
the $K_L$ decay length $L_K$ is obtained from an energy weighted average of 
the $L_{K,i}$ determined by each individual photon. The accuracy of this method
is tested in $K_L\rightarrow\pi^+\pi^-\pi^0$ events by comparing 
neutral and charged vertex positions, and its resolution is evaluated, using
the neutral decay events themselves, to be 
$\sim 2\, cm$ with a sligth dependence on the $K_L$ decay length. 
For this method to work it is crucial to correctly 
identify  the bunch crossing that originated the event.
This is done by identifying 
one of the two pions produced by the $K_S$ decay and by measuring its track
length $l_{\pi}$, momentum and time of flight of the associated calorimeter
cluster $t_{cl}$. By requiring $|l_{\pi}/\beta_{\pi} - t_{cl}|<2\, ns$ for at
least one of the $K_S$ pions, the probability of correct bunch crossing 
identification becomes (99.4$\pm$ 0.1)\%, as measured by comparing charged and
neutral vertex positions in a $K_L\rightarrow\pi^+\pi^-\pi^0$ sample.\par 
\begin{figure}[tb]
\begin{center}
\begin{tabular}{ll}
\epsfig{figure=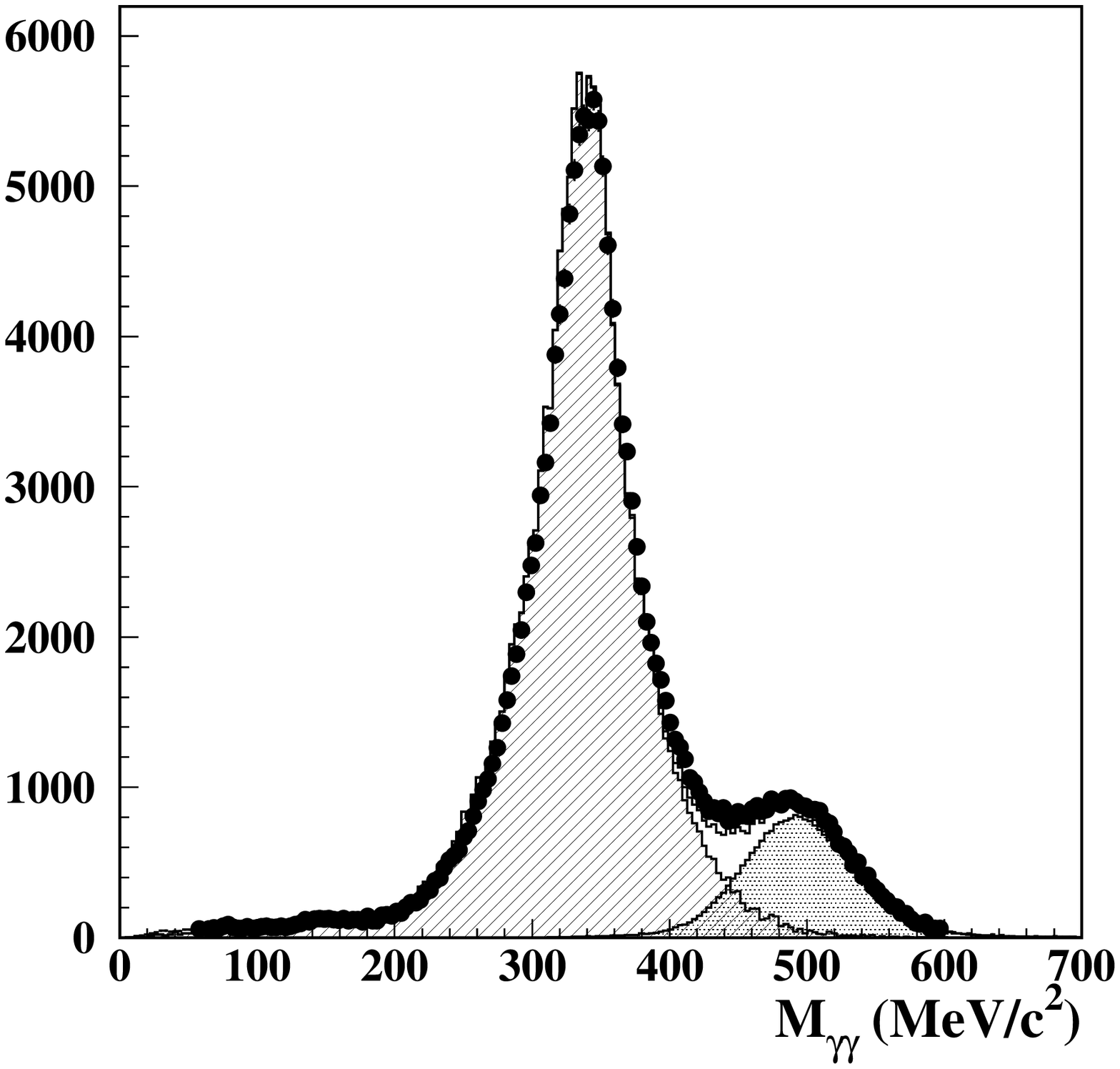,width=8cm} & \epsfig{figure=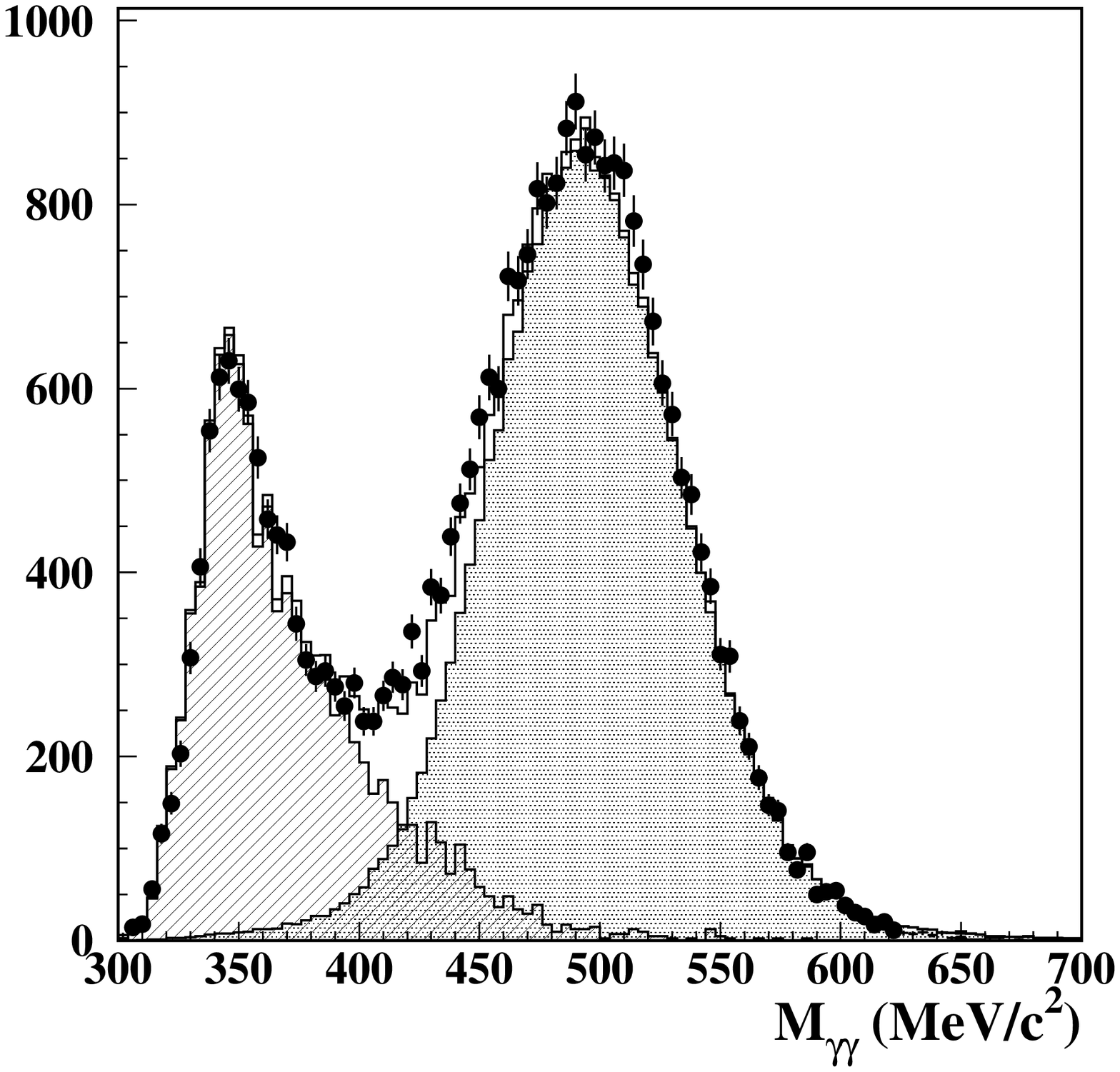,width=8cm} \\
\end{tabular}
\end{center}
\caption{$M_{\gamma\gamma}$ distribution before (left) and after (right) the 
$K_L\rightarrow\gamma\gamma$ final selection ($E^{\prime}$ and $\alpha$ cuts).
Dots are data, histograms are MC (full, signal and background).
\label{fig:kl2gam}}
\end{figure}
$K_L\rightarrow\pi^0\pi^0\pi^0$ decays are selected by requiring the presence 
of at least 3 calorimeter clusters not associated to any tracks, with energy
larger than $20\, MeV$, distance from any other cluster larger than $40\, cm$,
and yielding $L_K$ in the fiducial volume, with $|L_{K,i}-L_K|<4\sigma_{L_K}$.
The MC estimated selection efficiency is (99.80$\pm$ 0.01)\% . Only events with
3 and 4 clusters are contaminated by background, as it turns out from a data-MC
comparison of the total energy distributions. Such background is given by 
$K_L\rightarrow\pi^+\pi^-$ events and by misreconstructed $K_L\rightarrow\pi^+
\pi^-\pi^0$ decays. In order to reject it, tighter selections are applied
on energy and position of the clusters in the 3 and 4 clusters sample, and
events with tracks not associated to the $K_S$ decay and with
first hit in the drift chamber nearer than $30\, cm$ to the $K_L$ vertex are
vetoed. Finally the 3 and 4 clusters samples contribute only to less than 8\%
of the total statistics, with a  
residual background contamination, evaluated by MC, of 
(18.6$\pm$1.0)\% in the 3 clusters sample, and (7.0$\pm$0.2)\% in the 4 
clusters sample. The trigger efficiency, measured in data, is 
(99.88$\pm$0.04)\%. After background subtraction and efficiency correction, the
measured number of $K_L\rightarrow\pi^0\pi^0\pi^0$ decays in our data sample is
$N_{3\pi^0}=9,802,200\cdot (1\pm 0.0010 \pm 0.0016)$.\par
$K_L\rightarrow\gamma\gamma$ are first preselected by requiring the presence  
of at least 2 calorimeter clusters not associated to tracks, with energy $E> 
100\, MeV$. For the two most energetic such clusters, loose requirements are
made on total energy, relative angle, time difference and compatibility of 
$L_K$ determination.  The preselection efficiency is 89.5\% for signal events, 
evaluated by MC. In the left part of fig.\ref{fig:kl2gam} the distribution
of the invariant mass of the couples of clusters surviving the preselection is 
shown, together with the MC expectations for signal and background. At this 
stage a peak around $M_K$ is already present, but not yet resolved from 
background. To enhance it, the photon energies are computed with 
better accuracy assuming $E_{\gamma}=M_K/2$ in the $K_L$ center of mass and 
boosting them to the laboratory. The boost is obtained from $\vec{p_{K_L}}=
\vec{p_{\phi}}-\vec{p_{K_S}}$ and from clusters and $K_L$ decay vertex 
coordinates. The so computed total energy in the laboratory, $E^{\prime}$, is
required to be $|E^{\prime}-510|<5\sigma^{\prime}$, where $ \sigma^{\prime}=
1.8\, MeV$ was evaluated from a fit to the experimental $E^{\prime}$ 
distribution. Moreover the angle $\alpha$ between $\vec{p_{K_L}}$ and the
laboratory total $\gamma\gamma$ momentum (obtained from the recomputed photon
energies and their clusters positions) is required to be smaller than 
$15^{\circ}$. The final $M_{\gamma\gamma}$ distribution  is 
shown in the right part of fig.\ref{fig:kl2gam}. The selection efficiency was 
evaluated using a sample of $K_L\rightarrow\gamma\gamma$ with high purity
(S/B$\sim 10^3$) selected by applying hard, uncorrelated cuts on other 
kinematic variables\cite{kl2gam}, and turns out to be ($81.0\pm 0.3_{stat}\pm
0.5_{syst}$)\% . The trigger efficiency was evaluated from data as for $K_L
\rightarrow 3\pi^0$ decays. The number of $K_L\rightarrow\gamma\gamma$ decays 
was then obtained from a fit to the $M_{\gamma\gamma}$ distribution using MC
shapes for signal and background. After correcting for the efficiencies we get
$N_{2\gamma}=27,375\cdot (1\pm 0.0076 \pm 0.0081)$, which  yields
\begin{equation} 
\frac{\Gamma (K_L\rightarrow\gamma\gamma)}
     {\Gamma (K_L\rightarrow\pi^0\pi^0\pi^0)} = 
  (2.793\pm0.022_{stat}\pm 0.024_{syst})\cdot 10^{-3}
\label{eq:kl2gam} 
\end{equation} 
in good agreement with the recent NA48 measurement\cite{na48}. Using the known
value of BR($K_L\rightarrow\pi^0\pi^0\pi^0$) and $\tau_L$ we get $\Gamma (
K_L\rightarrow\gamma\gamma) = (7.5\pm0.1)\cdot 10^{-12}eV$, which is in 
agreement with the $o(p^6)$ predisctions of ChPT provided the value of the 
pseudoscalar mixing angle is close to our recent measurement\cite{eta}.\par
  
\section{Study of the decay $\phi\rightarrow\pi^+\pi^-\pi^0$}
The decay of the $\phi$ meson to $\pi^+\pi^-\pi^0$ is dominated by the 
$\rho\pi$ intermediate states $\rho^+\pi^-$, $\rho^-\pi^+$, $\rho^0\pi^0$ 
with equal amplitudes. Additional contributions to $e^+e^-\rightarrow\pi^+\
pi^-\pi^0$ are the so called ``direct term'' and $e^+e^-\rightarrow\omega\pi$
($\omega\rightarrow\pi^+\pi^-$). Taking into account all contributions, a fit
to the Dalitz plot of the process allows to determine masses and widths of the
three $\rho$ charge states\cite{rho}. CPT invariance requires equality of masses and 
widths of $\rho^+$ and $\rho^-$, while isospin-violating electromagnetic 
effects may originate possible differences in mass or width between $\rho^0$ 
and $\rho^{\pm}$. Starting from a 16 pb${}^{-1}$ data sample collected in 
year 2000, events are selected by requiring the presence of two non 
collinear tracks, with opposite sign curvature and polar angle $\theta 
>40^{\circ}$. The acollinearity cut is $\Delta\theta<175^{\circ}$, and rejects
$e^+e^-\gamma$ events without affecting the signal. Then imposing the 
conservation of the $\phi$ meson energy and momentum, the missing mass 
$M_{miss}$ is 
computed, and required to be within $20\, MeV$ of the $\pi^0$ mass. This 
corresponds to an effective cut of $<20\, MeV$ on the total ISR radiated 
energy. Finally two neutral clusters in the calorimeter are required, with
$E>10\, MeV$ and arrival time compatible with that of a photon emitted in the 
interaction point. The two photon opening angle in the $\pi^0$ rest frame must
have $cos\theta_{\gamma\gamma}<-0.98$. After such selection the residual
background contrinution is less tham $10^{-5}$, while the efficiency for the 
signal varies between 20\% and 30\% over the kinematic range of the
process.\par 

\begin{table}[htb]
\caption{Fitted results for $\rho$ masses and widths (in MeV).
 \label{tab:rhopi}}
\vspace{0.4cm}
\begin{center}
\begin{tabular}{|c|c|c|c|}
\hline
parameter & fit(a) & fit(b) & fit(c)\\  
\hline
$M_{\rho^0}$ & & $775.9\pm0.5\pm 0.5$ & $775.9\pm0.6\pm 0.5$ \\
\cline{3-4}
$M_{\rho^+}$ & $775.5\pm0.5\pm 0.3$  & $775.5\pm0.5\pm 0.4$ & $776.3\pm0.6\pm 0.7$ \\
\cline{4-4}
$M_{\rho^-}$ & & & $774.8\pm0.6\pm 0.4$ \\
\hline
$\Gamma_{\rho^0}$ & & $147.3\pm1.5\pm 0.7$ & $147.4\pm1.5\pm 0.7$ \\
\cline{3-4}
$\Gamma_{\rho^+}$ & $143.9\pm1.3\pm 1.1$ & $143.7\pm1.3\pm 1.2$ & $144.7\pm1.4\pm 1.2$ \\
\cline{4-4}
$\Gamma_{\rho^-}$ & &  & $142.9\pm1.3\pm 1.4$ \\
\hline
\end{tabular}
\end{center}
\end{table}

The Dalitz plot is built up as a function of the two variables $x=T^+ - T^-$ 
and $y=T^0$, where $T^{+-0}$ are the kinetic energies of $pi^{+-0}$ in the
center of mass system. The resolution on $x$ and $y$ is about $1\, MeV$ over
the full kinematic range. The Dalitz plot is divided in $8.75\times 8.75\,
MeV^2$ bins, for a total of 1874 bins within the kinematic boundary. The 
trigger and selection efficiency have been evaluated as a function of  $x$ and 
$y$ using MC simulation with corrections based on data control samples. In
particular the the low energy photon detection efficiency has been obtained
from $e^+e^-\gamma$ events. The calibration of the momentum scale is checked
across the entire kinematic range by comparing the measured $M_{miss}$
 with the $\pi^0$ mass. Three fits have been performed to the Dalitz plot 
density distribution: a) a fit assuming CPT and isospin invariance, {\it i.e.}
$M_{\rho^0}=M_{\rho^+}=M_{\rho^-}$ and $\Gamma_{\rho^0}=\Gamma_{\rho^+}=
\Gamma_{\rho^-}$; b) a fit assuming only CPT invariance, {\it i.e.} 
$M_{\rho^+}=M_{\rho^-}$ and $\Gamma_{\rho^+}=\Gamma_{\rho^-}$; c) a fit without
limitations on masses and widths. The fit results are shown in table 
\ref{tab:rhopi}. Systematic uncertainties come mainly from the fit stability 
with respect to the selection cuts and from the absolute momentum calibration.
The $\rho$ masses are significantly larger and the widths smaller than the 
PDG averages\cite{pdg}, but are close to the most recent measurements. 
The direct term and $\omega\pi$ contributions were also included in the fits,
and found to be significantly different from zero. In particular the visible
cross section for the process $e^+e^-\rightarrow\omega\pi^0\rightarrow\pi^+
\pi^-\pi^0$ at $\sqrt{s}=1019.4\, MeV$ turns out to be $92\pm15\, pb$.\par 


\section{$\phi$ radiative decays}
The $\phi$ meson radiative decays are studied in details at KLOE. Three main 
measurements have been performed with the year 2000 data sample: the 
BR($\phi\rightarrow\eta^{\prime}\gamma$)\cite{eta}, the {BR($\phi\rightarrow 
f_0\gamma$)\cite{f0} and the BR($\phi\rightarrow a_0\gamma$)\cite{a0}. All
such measurements are being updated with more statistics.\par
The $\phi\rightarrow\eta^{\prime}\gamma$ decay is identified in the channel in 
which $\eta^{\prime}\rightarrow\eta\pi^+\pi^-$ and $\eta\rightarrow\gamma
\gamma$. The $\phi\rightarrow\eta\gamma$ decay is identified in the channel in 
which $\eta\rightarrow\pi^+\pi^-\pi^0$. In both cases the final state is $\pi^+
\pi^-\gamma\gamma\gamma$, so that many systematics cancel in the ratio
$BR(\phi\rightarrow\eta^{\prime}\gamma/BR(\phi\rightarrow\eta\gamma)$. Events 
with two oppositely charged tracks and three prompt clusters are selected.
Background from $\phi\rightarrow K_SK_L$ and $\phi\rightarrow\pi^+\pi^-\pi^0$
are rejected by simple cuts on particles' energy and momentum. $\eta^{\prime}
\gamma$ and $\eta\gamma$ decays are disentangled by a cut on the $E_1 ,E_2$
plane, the energies of the two most energetic photons. The number of  
$\eta^{\prime}\gamma$ events is then obtained from a fit to the reconstructed 
$\eta^{\prime}$ invariant mass: in the year 2000 sample we have $124\pm 12\pm
 5$ events in the peak, and in the 2001 sample we have already more than 700
events in the peak. The published\cite{eta} result $BR(\phi\rightarrow\eta^{\prime}\gamma
/BR(\phi\rightarrow\eta\gamma)=(4.70\pm0.47\pm0.31)\cdot 10^{-3}$ corresponds 
to a pseudoscalar mixing angle: in the flavour basis $\varphi_P=
(41.8^{+1.9}_{-1.6})^{\circ}$, and in the singlet-octet basis 
$\vartheta_P=(12.9^{+1.9}_{-1.6})^{\circ}$. Using the PDG value for 
$BR(\phi\rightarrow\eta\gamma)$ we get $BR(\phi\rightarrow\eta^{\prime}\gamma=
(6.10\pm0.61\pm0.43)\cdot 10^{-5}$, which is compatible with zero gluonium
contribution to the $\eta^{\prime}$ state.\par
The $\phi\rightarrow f_0\gamma$ and $\phi\rightarrow a_0\gamma$ are recognized
at KLOE by their five photons final states ($f_0\rightarrow\pi^0\pi^0$ and
$a_0\rightarrow\eta\pi^0$, $\eta\rightarrow\gamma\gamma$). Various backgrounds
are present, the main ones coming from $\phi\rightarrow\rho^0\pi^0$ and from
$e^+e^-\rightarrow\omega\pi^0$, which are rejected by means of kinematic fits
asumin various event topologies. Cuts are applied to the reconstructed masses 
of the intermediate particles and on the $\chi^2$ probability of the fits. 
Published KLOE measurements are $BR(\phi\rightarrow\pi^0\pi^0\gamma)=(1.09\pm
0.03\pm0.05)\cdot 10^{-4}$ and  $BR(\phi\rightarrow\eta\pi^0\gamma)=(1.08\pm
0.05\pm0.06)\cdot 10^{-4}$. Higher statistics analyses are being performed
and are in agreement with the present results. Finally, the mass spectra of 
the two processes are sensitive to the nature of $f_0$ and $a_0$: the spectra
measured by KLOE are compatible with a $qq\bar q \bar q$ nature for the $f_0$,
while for the $a_0$ conclusions cannot be drawn up to now.\par

\section{Hadronic cross section}
The recent measurements of $a_{\mu}$ by the E821 collaboration\cite{e821}
has stimulated new interest in the measurement of the cross section of 
$e^+e^-\rightarrow hadrons$ at low energy. In fact, the hadronic contribution 
to $a_{\mu}$ at low energy cannot be computed, but is related to $\sigma
(e^+e^-\rightarrow hadrons)$ via a dispersion integral. The process $e^+e^-
\rightarrow\pi^+\pi^-$, with $M_{\pi\pi}<1\, GeV$ accounts for about 70\%
of $\delta a_{\mu}^{had}$ and for 12\% of the hadronic corrections to
$\alpha (M_Z)$. KLOE can measure $d\sigma(e^+e^-\rightarrow\pi^+
\pi^- /dM^2_{\pi\pi}$ by studying $e^+e^-\rightarrow\pi^+\pi^-\gamma$ events
in which the photon is radiated in the initial state. This has the advantage
that sistematic erors due to luminosity and beam energy enter into the 
measurement only once, and not for each point as in an energy scan, but on the
other hand it requires theoretic understanding of ISR to better than 1\% . 
Events are selected requiring two oppositely charged
tracks, with polar angle $\theta <40^{\circ}$ and
forming a vertex near the interaction region. To enhance ISR, with respect
to FSR and $\phi\rightarrow\pi^+\pi^-\pi^0$ background, only small angle 
photons are accepted. As the forward and backward angles within a cone of
$15^{\circ}$ are obscured by the QCALs,
photons are not required to be detected, but the $\pi^+\pi^-$ missing momentum
is required to have polar angle < $15^{\circ}$. Moreover at least one of the 
two tracks has to be identified as a pion, on the basis of time of flight and
shower shape in the calorimeter. Selection efficiency is better than 96\%.
Using a 73 pb${}^{-1}$ data sample, collected in year 2001, KLOE has now a
 $d\sigma(e^+e^-\rightarrow\pi^+\pi^- /dM^2_{\pi\pi}$ distribution, for
$M^2_{\pi\pi}$ ranging from 0.2 to 1 $GeV^2$, with statistical error $\sim 1$
in each bin. In the coming months KLOE will double the statistics and reduce
the systematic errors to below the 1\% level.\par


\end{document}